\documentclass{article}
\usepackage{spconf,amsmath,amssymb,graphicx,hyperref,xcolor}
\usepackage{booktabs}
\usepackage{multirow}
\usepackage{xcolor}
\definecolor{darkgreen}{RGB}{0,128,0}
\definecolor{darkred}{RGB}{150,0,0}
\usepackage{enumitem}
\usepackage{colortbl}   
\usepackage[skip=-1pt]{caption}
\usepackage{subcaption}
\usepackage{lineno}
\usepackage{setspace}

\title{Endocaver: Handling Fog, Blur and Glare in Endoscopic Images via Joint Deblurring-Segmentation}
\name{Zhuoyu Wu$^{1}$, Wenhui Ou$^{2}$, Pei-Sze Tan$^{1}$, Jiayan Yang$^{3}$, Wenqi Fang$^{3}$, Zheng Wang$^{3}$, Rapha\"{e}l C.-W. Phan$^{1\dag}$
\thanks{$\dag$ Corresponding author: \texttt{raphael.phan@monash.edu} 
}
}

\address{$^1$CyPhi ($\Psi\Phi$) AI Research Lab, School of IT, Monash University, Malaysia Campus; 
\\$^2$Department of Electronic \& Computer Engineering, The Hong Kong University of Science and Technology; 
\\$^3$Shenzhen Institute of Advanced Technology, Chinese Academy of Sciences;}
\begin{document}
%
\maketitle
\begin{abstract}
Endoscopic image analysis is vital for colorectal cancer screening, yet real-world conditions often suffer from lens fogging, motion blur, and specular highlights, which severely compromise automated polyp detection. We propose \textbf{EndoCaver}, a lightweight transformer with a unidirectional-guided dual-decoder architecture, enabling joint multi-task capability for image deblurring and segmentation while significantly reducing computational complexity and model parameters. Specifically, it integrates a Global Attention Module (GAM) for cross-scale aggr   egation, a Deblurring-Segmentation Aligner (DSA) to transfer restoration cues, and a cosine-based scheduler (LoCoS) for stable multi-task optimisation. Experiments on the Kvasir-SEG dataset show that EndoCaver achieves 0.922 Dice on clean data and 0.889 under severe image degradation, surpassing state-of-the-art methods while reducing model parameters by 90\%.
These results demonstrate its efficiency and robustness, making it well-suited for on-device clinical deployment. Code is available at \url{https://github.com/ReaganWu/EndoCaver}.
\end{abstract}

\section{Introduction}
\label{sec:Intro}
Colorectal cancer is the third most common cancer worldwide, accounting for nearly 10\% of all cancer cases, and the second leading cause of cancer-related deaths globally~\cite{morgan2023global}. Early detection of colorectal polyps through endoscopy is an effective preventive strategy. However, real-world endoscopic imaging often suffers from severe quality degradations such as lens fogging, motion blur, and specular highlights~\cite{munzer2013noisysource}. These degradations not only compromise the accuracy of automated polyp detection but may also mislead clinical diagnosis~\cite{nazarian2021diagnostic}.

\begin{figure}[ht]
    \centering
    \includegraphics[width=0.5\textwidth]{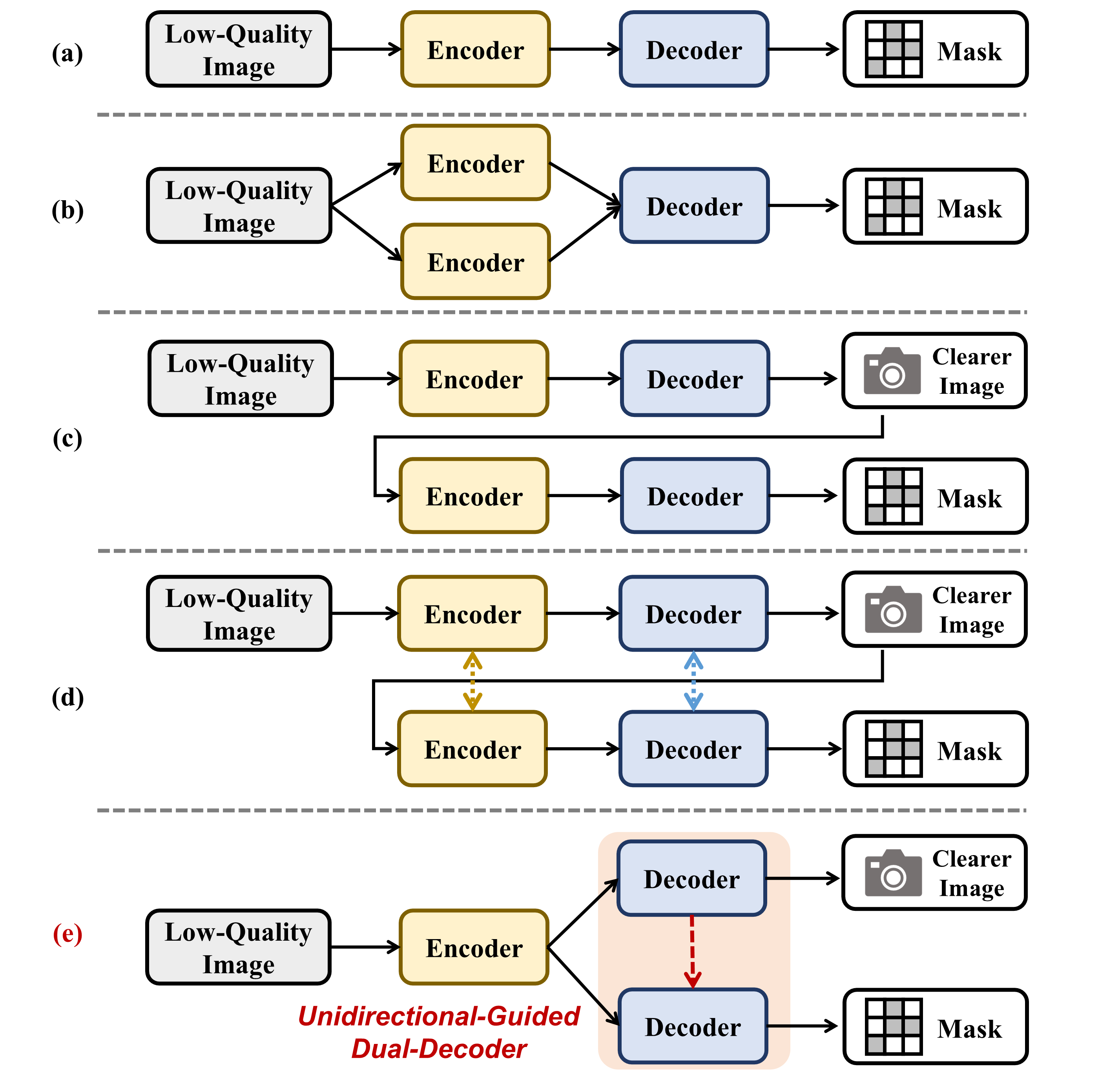}
    \caption{Comparison of different model designs. (a–d) illustrate representative existing forms, while (e) shows our proposed Unidirectional-Guided Dual-Decoder, the core of our work. It reduces overall model parameters and computational cost while enabling higher segmentation accuracy.}
    
    \label{fig:intro_pic}
\end{figure}

With the advances of deep learning in medical imaging, several efforts have been made to address this issue (Fig.~\ref{fig:intro_pic}). Traditional Single-task segmentation networks like UNet~\cite{ronneberger2015unet} and Segformer~\cite{xie2021segformer}(Fig.~\ref{fig:intro_pic}(a)) achieve reasonable performance primarily through improved backbone and fusion module architectures, yet these models' parameter efficiency remains suboptimal \cite{wu2024harmonizing}. Multi-encoder fusion methods~\cite{li2025cfformer}(Fig.~\ref{fig:intro_pic}(b)) improve robustness by using extra encoder(s) to extract information to leverage robustness but incur substantial computational costs that hinder real-time deployment. 
Two-stage pipelines (e.g., DRPViT~\cite{li2024drpvit}, Fig.~\ref{fig:intro_pic}(c)) decouple enhancement and segmentation, which increases training costs and prevents effective cross-task feature sharing. 
Joint-learning frameworks~\cite{wu2024harmonizing, dai2024i2u} (Fig.~\ref{fig:intro_pic}(d)) alleviate this problem by training enhancement and segmentation simultaneously, though at the cost of increased architectural complexity due to complicated information sharing between components. 
Critically, these existing solutions primarily rely on large-scale architectures, which are impractical for clinical deployments where computational resources are limited and real-time processing is required. 


To address these challenges, we propose \textbf{EndoCaver}, a lightweight dual-decoder transformer for joint endoscopic image deblurring and segmentation (Fig.~\ref{fig:intro_pic}(e)). EndoCaver introduces a Global Attention Module (GAM) for cross-scale feature enhancement, a Deblurring-Segmentation Aligner (DSA) for cross-task guidance, and a cosine annealing-based loss scheduler (LoCoS) for adaptive multi-task optimization.

Our main contributions are:
\begin{itemize}
    \vspace{-1mm}
    \item We design \textbf{EndoCaver}, a 7.8M-parameter dual-decoder model that achieves competitive deblurring and segmentation with only 11.9 GMACs, suitable for real-time deployment.
    \vspace{-1mm}
    \item We propose \textbf{GAM} and \textbf{DSA} to enhance encoder features and transfer restoration priors to segmentation for finer structural detail. 
    \vspace{-1mm}
    \item We introduce \textbf{LoCoS}, a cosine-annealing multi-task loss weighting strategy that accelerates convergence and improves out-of-distribution generalization.
\end{itemize}



\vspace{-2mm}
\section{Methodology}
\label{sec:method}

\begin{figure*}[ht]
    \centering
    \includegraphics[width=0.95\textwidth]{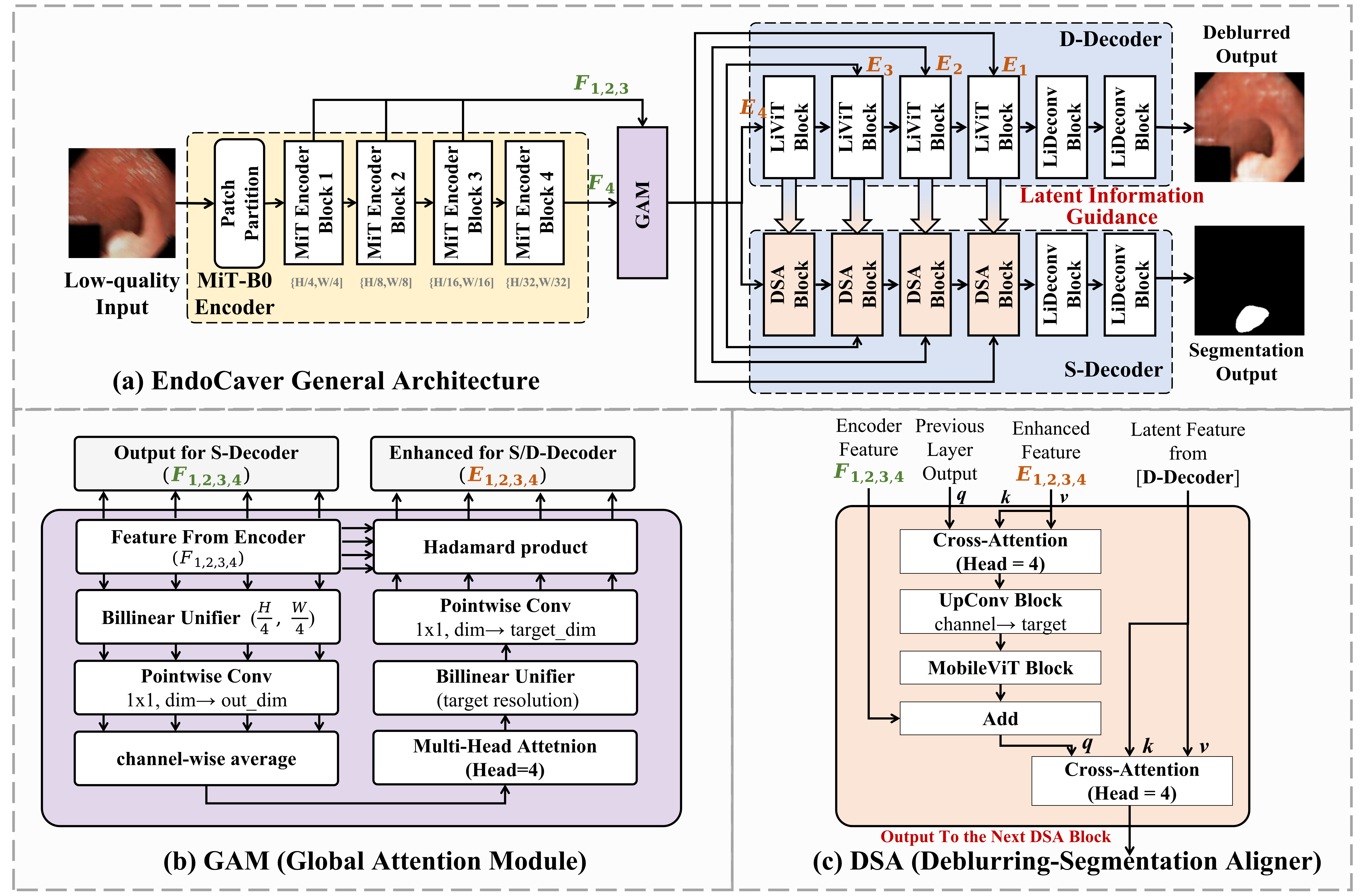}
    \caption{Overall architecture of EndoCaver. (a) End-to-end framework with MiT encoder, GAM, and dual decoders. (b) Global Attention Module (GAM) for lightweight cross-scale aggregation. (c) Deblurring-Segmentation Aligner (DSA) for cross-task feature alignment. 
    For decoder blocks, LiDeconv employs transposed convolution with BatchNorm and SiLU for recovery, followed by a reverse residual bottleneck~\cite{sandler2018mobilenetv2} for feature extraction. LiViT only replaces the bottleneck with MobileViT~\cite{mehta2021mobilevit}.}
    \label{fig:Architecture}
\end{figure*}

\subsection{Overall Architecture}
EndoCaver is a lightweight dual-decoder transformer designed for joint \textbf{endoscopic image deblurring} and \textbf{polyp segmentation} under real-world degradations. As shown in Fig.~\ref{fig:Architecture}(a), the framework consists of:  
(i) an \textbf{MiT-B0 encoder}\cite{xie2021segformer} for efficient hierarchical representation,  
(ii) a \textbf{Global Attention Module (GAM)} for cross-scale aggregation, and  
(iii) a \textbf{unidirectional dual-decoder structure} with Deblurring-Segmentation Aligner (DSA) to transfer deblurring priors into segmentation.  

Formally, given an input $I \in \mathbb{R}^{H \times W \times 3}$, the encoder produces multi-scale features:
\begin{equation}
\{F_1, F_2, F_3, F_4\} = \text{Encoder}(I),
\end{equation}
where $F_i \in \mathbb{R}^{C_i \times H_i \times W_i}$ are hierarchical features.  
The D-Decoder restores the deblurred image $\hat{I}$, while the S-Decoder predicts the segmentation mask $\hat{M}$:
\begin{equation}
\begin{aligned}
\hat{I} &= \text{D-Decoder}(F_1,\dots,F_4), \\
\hat{M} &= \text{S-Decoder}(F_1,\dots,F_4, \hat{I}).
\end{aligned}
\end{equation}

\vspace{-3mm}
\subsection{Global Attention Module}
Traditional U-Net skip connections or MIMO-Unet style fusion incur high complexity. To enable efficient cross-scale fusion, we unify encoder features to a fixed resolution $[H/4, W/4]$:
\begin{equation}
F'_i = \text{Resize}(F_i) \in \mathbb{R}^{128 \times H/4 \times W/4}.
\end{equation}
Channel-wise averaging reduces redundancy:
\begin{equation}
F_{\text{avg}} = \frac{1}{N}\sum_{i=1}^{N} F'_i.
\end{equation}
Then, multi-head attention (MHA) enhances global relations:
\begin{equation}
\text{MHA}(Q,K,V) = \text{Softmax}\!\left(\frac{QK^\top}{\sqrt{d}}\right)V,
\end{equation}
where $Q,K,V$ are linear projections of $F_{\text{avg}}$.  
The enhanced feature is fused back through bilinear upsampling and pointwise convolution via 
Hadamard 
product with the original input to produce Enhanced Feature $E_{\text{1,2,3,4}}$ for decoders.  
\vspace{-3mm}
\subsection{Deblurring-Segmentation Aligner}
While existing decoders process segmentation independently, EndoCaver bridges tasks via cross-attention \cite{radford2021clip, vaswani2017attention} (Fig.~\ref{fig:Architecture}(c)).  
Given segmentation query feature $F_s$ and latent deblurring feature $F_d$, the DSA performs two stages:  

\noindent
\textbf{Stage 1:} Align segmentation with encoder-enhanced features:
\begin{equation}
F_s' = \text{CrossAttn}(F_s, F_{\text{enc}}),
\end{equation}
where $F_{\text{enc}}$ is from the GAM output.  

\noindent
\textbf{Stage 2:} Inject latent deblurring priors:
\begin{equation}
F_s'' = \text{CrossAttn}((F_s'+F_i), F_d).
\end{equation}
Where $F_i$ is from the Encoder features.

The $F_s'$ is further refined by a MobileViT block \cite{mehta2021mobilevit} and upsampled to yield final segmentation logits. Then, added with the residual encoder feature $F_i$ to preserve information. The last part of GAM is cross-attention with the deblurring feature $F_d$, to obtain 
the deblurring-guided feature $F_s''$.

\vspace{-3mm}
\subsection{Loss with Cosine Annealing (LoCoS)}
We jointly optimize deblurring and segmentation with task-adaptive weighting. The total loss is
\begin{equation}
\mathcal{L}(t) = \big(1-w_{\text{seg}}(t)\big)\mathcal{L}_{\text{deb}} + w_{\text{seg}}(t)\mathcal{L}_{\text{seg}},
\end{equation}
where
\begin{equation}
\mathcal{L}_{\text{deb}} = \| I - \hat{I} \|_2^2, 
\quad \mathcal{L}_{\text{seg}} = 1 - \text{Dice}(M, \hat{M}),
\end{equation}
and $w_{\text{seg}}(t)$ follows a cosine annealing schedule:
\begin{equation}
w_{\text{seg}}(t) = w_{\min} + \tfrac{1}{2}(1-w_{\min})(1+\cos(\pi t / T)).
\end{equation}
This gradually shifts focus from deblurring to segmentation, ensuring stable joint optimization.

\vspace{-1mm}
\section{Experiments and Results}
\label{sec:exp}
\begin{table}[t]
\scriptsize
\centering
\caption{Performance comparison on Kvasir dataset under clean and degraded (noisy) conditions. Parameters (M) and complexity (GMac) are also reported. ($\uparrow$) / ($\downarrow$) indicate higher / lower is better.}
\begin{tabular}{lcccc}
\hline
Method & Clean ($\uparrow$) & Degraded ($\uparrow$) & \#Params ($\downarrow$)& GMac \\
\hline
UNet\cite{ronneberger2015unet}              & 0.8699 & 0.8118 (\textcolor{darkgreen}{-0.0581}
) & 31.04 & 41.93 \\
I2UNet-L\cite{dai2024i2u}          & 0.8556 & 0.7982 (\textcolor{darkgreen}{-0.0574}
)& 29.65 & 9.35 \\
CTNet\cite{xiao2024ctnet}             & 0.8855 & 0.8270 (\textcolor{darkgreen}{-0.0585}
)& 44.29 & 6.27 \\
NPDnet\cite{yu2024novel}            & 0.8945 & 0.8286 (\textcolor{darkgreen}{-0.0659}
)& 27.67 & 5.14 \\
CFFormer\cite{li2025cfformer}          & 0.9127 & 0.8507 (\textcolor{darkgreen}{-0.0620}
)& 99.56 & 30.12 \\
Segformer-B5\cite{xie2021segformer}      & 0.9195 & 0.8623 (\textcolor{darkgreen}{-0.0572}
)& 81.97 & 12.35 \\
\hline
H-Unets\cite{wu2024harmonizing}           & 0.9011 & 0.8325 (\textcolor{darkgreen}{-0.0686}
)& 16.23 & 12.78 \\
\rowcolor{gray!15}\textbf{EndoCaver} & \textbf{0.9221} & \textbf{0.8893} (\textcolor{darkred}{-0.0328}
)& \textbf{7.81} & \textbf{11.86} \\
\hline
\label{tab:analysisdegrade}
\end{tabular}
\end{table}

\subsection{Experimental Setup}
Our model is implemented in PyTorch and trained on a single NVIDIA A100 80G GPU. Input images are resized to $224 \times 224$ with a batch size of 16. Training is with the Adam optimizer, warmup, and cosine annealing learning rate schedule from $1 \times 10^{-4}$ during epochs (Deblurring, Endocaver:3000, Segmentation:200). As described in Section~\ref{sec:method}, the LoCoS strategy is used to balance Dice and MSE losses.

\vspace{-2mm}
\subsection{Datasets and Metrics}
Following standard protocols, we use 900 images from Kvasir-SEG \cite{jha2019kvasir} (80\% training, 20\% validation) and evaluate out-of-distribution generalization on CVC-ClinicDB \cite{bernal2015cvcclinic} and CVC-ColonDB \cite{bernal2012colondb}. Segmentation is assessed by Dice, IoU, and Recall, while deblurring quality is measured by PSNR and SSIM (higher is better).

\textit{Synthetic Degradations.}
To evaluate robustness under adverse imaging conditions, we generate degraded images with motion/defocus blur, specular highlights, and lens fogging using randomly sampled parameters. This complements clean datasets and enables systematic stress testing.
\vspace{-3mm}

\begin{table*}[ht]
\centering
\caption{Performance comparison of different methods on noisy datasets. Metrics include Dice coefficient, IoU, Recall, model parameters, computational complexity, and model type.}
\label{tab:results_noisy}
\scriptsize
\begin{tabular}{l|ccc|cc|c}
\toprule
\multirow{3}{*}{\textbf{Method}} & 
\multicolumn{3}{c|}{\textbf{Dataset Performance}} & 
\multicolumn{2}{c|}{\textbf{Model Efficiency}} &
\multirow{3}{*}{\textbf{Type}} \\
\cmidrule(lr){2-4} \cmidrule(lr){5-6}
& \textbf{Kvasir} & \textbf{Kvasir} $\boldsymbol{\rightarrow}$ \textbf{ClinicDB} & \textbf{Kvasir} $\boldsymbol{\rightarrow}$ \textbf{ColonDB} & \textbf{Params} & \textbf{Complexity} & \\
& \textbf{(Noisy, Trained)} & \textbf{(Noisy, OOD Test)} & \textbf{(Noisy, OOD Test)} & \textbf{(M)} & \textbf{(GMac)} & \\
\cmidrule(lr){2-4} \cmidrule(lr){5-6}
& Dice / IoU / Recall & Dice / IoU / Recall & Dice / IoU / Recall & & & \\
\midrule
\multicolumn{7}{c}{\textit{Segmentation Only Methods}} \\
\midrule
UNet\cite{ronneberger2015unet}      & 0.812 / 0.683 / 0.807 & 0.634 / 0.472 / 0.764 & 0.500 / 0.339 / 0.681 & 31.04 & 41.93 & A \\
I2UNet-L\cite{dai2024i2u}  & 0.798 / 0.661 / 0.808 & 0.553 / 0.387 / 0.727 & 0.494 / 0.331 / 0.764 & 29.65 & 9.35  & D \\
CTNet\cite{xiao2024ctnet}     & 0.827 / 0.696 / 0.795 & 0.669 / 0.507 / 0.732 & 0.573 / 0.407 / 0.763 & 44.29 & 6.27  & A \\
NPDNet\cite{yu2024novel}    & 0.829 / 0.698 / 0.793 & 0.676 / 0.515 / 0.731 & 0.560 / 0.392 / 0.682 & 27.67 & 5.14  & C \\
CFFormer\cite{li2025cfformer}  & 0.851 / 0.744 / 0.858 & 0.728 / 0.575 / 0.777 & 0.616 / 0.449 / 0.735 & 99.56 & 30.12 & B \\
SegFormer-B5 & 0.862 / 0.762 / 0.874 & \textbf{0.782} / 0.648 / \textbf{0.764} & 0.685 / 0.529 / 0.718 & 81.97 & 12.35 & A \\
\midrule
\multicolumn{7}{c}{\textit{Deblurring + Segmentation Methods}} \\
\midrule
H-UNets\cite{wu2024harmonizing}   & 0.833 / 0.713 / 0.801 & 0.673 / 0.509 / 0.736 & 0.572 / 0.410 / 0.702 & 16.23 & 12.78 & D \\
DRPVit\cite{yu2024novel} & 0.845 / 0.735 / 0.848 & 0.741 / 0.592 / 0.753 & 0.689 / 0.539 / 0.773 & 87.21 & 34.31 & C \\
Cascaded$^1$ & 0.840 / 0.731 / 0.845 & 0.750 / 0.611 / 0.751 & 0.672 / 0.508 / 0.673 & 21.34 & 120.03 & C \\
Cascaded$^2$ & 0.851 / 0.745 / 0.858 & 0.736 / 0.587 / 0.702 & 0.665 / 0.501 / 0.713 & 11.08 & 13.68 & C \\
\rowcolor{gray!15}
\textbf{EndoCaver} & \textbf{0.889} / \textbf{0.803} / \textbf{0.872} & \textbf{0.782} / 0.644 / \textbf{0.764} & \textbf{0.702} / \textbf{0.552} / 0.711 & \textbf{7.81} & \textbf{11.86} & E \\
\bottomrule
\end{tabular}
\vspace{0.4em}
\begin{minipage}{\textwidth}
\footnotesize
\textbf{Notes:} Best results are in \textbf{bold}. 
\emph{Noisy, OOD Test}: unseen noisy datasets, trained on Kvasir (Noisy).
Cascaded$^1$ = MIMOUnetPlus~\cite{cho2021rethinking} + EndoCaver (S-Decoder);
Cascaded$^2$ = RT-Focuser~\cite{wu2025rt} + EndoCaver (S-Decoder).
PSNR / SSIM: H-UNet 21.12 / 0.9115, DRPViT 24.28 / 0.9431, 
Cascaded$^1$ 25.27 / 0.9654, Cascaded$^2$ 24.52 / 0.9602.
Types A--E $\rightarrow$ Enc--Dec, Dual-Enc, Cascaded, Fusion-Cascaded, Proposed Dual-Dec, corresponding to Fig.~\ref{fig:intro_pic} (a–e).
\end{minipage}
\end{table*}

\vspace{-2mm}
\subsection{Results and Discussions}
\vspace{-2mm}

To ensure fair evaluation on both target and unseen datasets, we compare EndoCaver with representative methods from all categories in Fig.~\ref{fig:intro_pic}:
\begin{itemize}[itemsep=1pt, topsep=2pt, parsep=0pt, partopsep=0pt]
    \item Type a (Encoder–Decoder): UNet~\cite{ronneberger2015unet}, Segformer-B5~\cite{xie2021segformer}, CTNet~\cite{xiao2024ctnet}
    \item Type b (Dual-Encoder): CFFormer~\cite{li2025cfformer}
    \item Type c (Cascaded): NPDnet~\cite{yu2024novel}, DRPVit~\cite{li2024drpvit}, and cascaded variants (MIMOUnet~\cite{cho2021rethinking}+EndoCaver(S-Decoder))
    \item Type d (Fusion-Cascaded): H-Unets~\cite{wu2024harmonizing}, I2UNet-L~\cite{dai2024i2u}
\end{itemize}


\noindent
\textbf{Degradation Analysis:} Table~\ref{tab:analysisdegrade} shows all methods experience performance drops under degraded conditions (-0.057 to -0.069), while EndoCaver achieves the highest clean Dice score (0.9221) with minimal degradation (-0.0320, ~50\% lower than competitors). Crucially, EndoCaver uses only 7.81M parameters and 11.86 GMac versus heavy models like CFFormer (99.56M) and Segformer-B5 (81.97M).

\begin{table}[t]
\centering
\scriptsize
\caption{Ablation study on Kvasir (Noisy).}
\label{tab:ablation}
\begin{tabular}{lcccc}
\toprule
Variant & Dice & PSNR & Params (M) & GMac \\
\midrule
\textbf{Full(EndoCaver)} & 0.8893 & 23.10 & 7.81 & 11.86 \\
\midrule
w/o LoCoS   & 0.8822 & 22.35 & 7.81 & 11.86 \\
w/o LoCoS\&DSA   & 0.8801 & 22.17 & 7.28 & 9.91 \\
w/o LoCoS\&DSA\&GAM   & 0.8756 & 23.06 & 7.08 & 4.36 \\
w/o Deblurring Branch & 0.8232 & -- & 5.23 & 1.68\\
\bottomrule
\end{tabular}
\end{table}

\noindent
\textbf{Quantitative Evaluation:}
As shown in Table~\ref{tab:results_noisy}, EndoCaver achieves superior results across datasets: Dice scores of 0.889(Kvasir), 0.782(ClinicDB), and 0.702(ColonDB), the generalization ability outperforming and matching advanced segmentation-only methods like SegFormer-B5, CFFormer, and other deblurring-segmentation methods.

\begin{figure}[ht]

    \hspace*{0.03\textwidth}
    \includegraphics[width=0.33\textwidth]{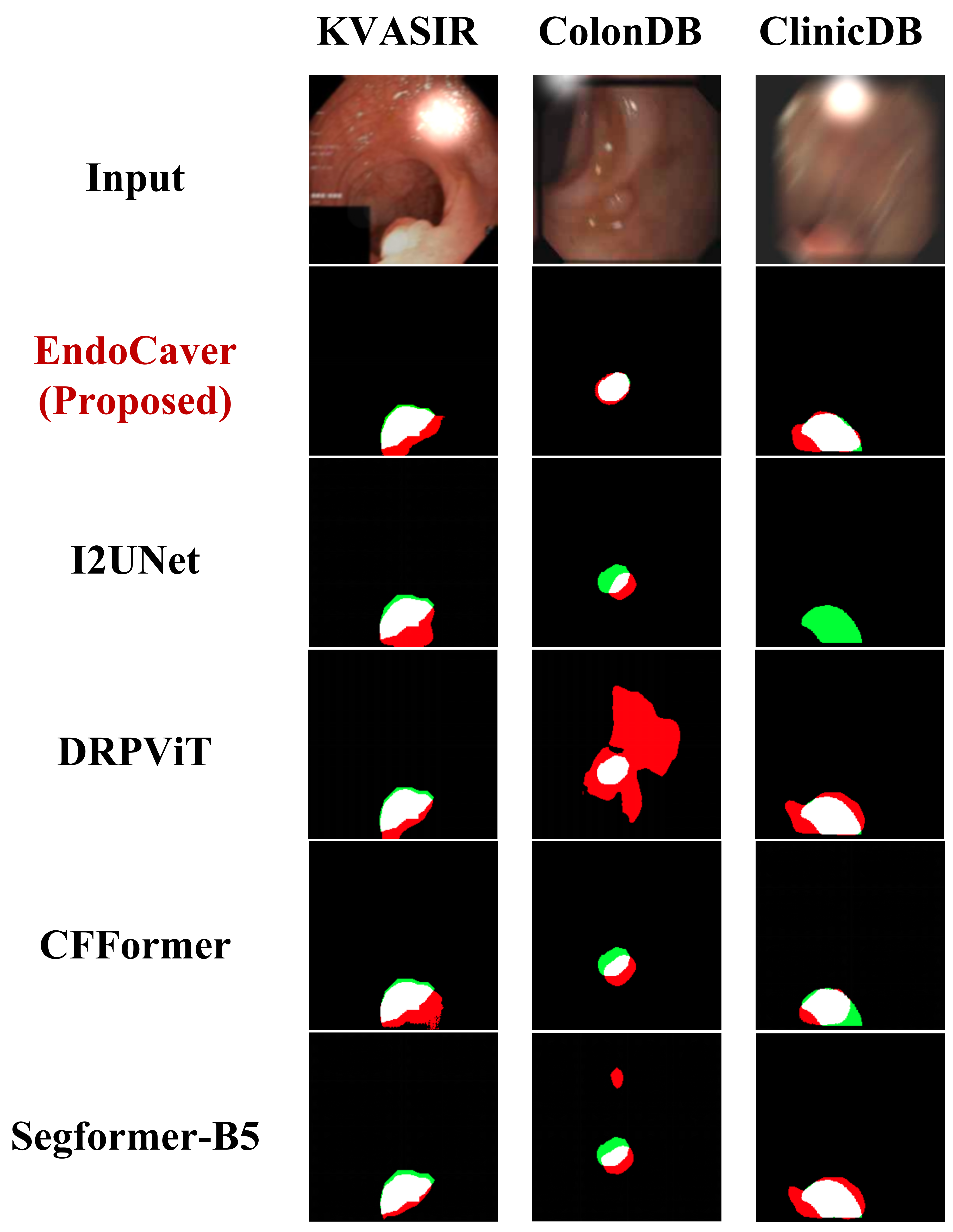}
    \caption{The qualitative results. Our proposed EndoCaver is labelled in red. The following model corresponds to the existing model design from Fig.\ref{fig:intro_pic} (d) to (a).}
    \label{fig:qualitative_results}
\end{figure}

\noindent
\textbf{Qualitative Evaluation:}
As illustrated in Fig.~\ref{fig:qualitative_results}, red regions indicate false segmentation, while green regions highlight missed segmentation. Compared to other methods, our approach delivers the most consistent results, particularly on unseen datasets (ColonDB and ClinicDB, columns two and three). This owing to its ability to restore structural details and guide segmentation.

\noindent 

\textbf{Ablation Studies:} Table~\ref{tab:ablation} shows the ablation results on Kvasir (Noisy). Removing LoCos slightly decreases Dice (0.8893 → 0.8822) and PSNR (23.10 → 22.35), indicating its role in stabilising multi-task learning. Further discarding DSA leads to a larger drop in Dice (0.8801) while also reducing computational cost. When all three components (LoCos, DSA, GAM) are removed, performance further degrades (0.8756 Dice), confirming that each component contributes to the overall effectiveness of EndoCaver.  
Notably, the segmentation-only baseline (w/o Deblurring Branch) achieves only 0.8232 Dice, highlighting that the strong performance of EndoCaver fundamentally relies on the deblurring branch and the feature-level synergy between deblurring and segmentation, rather than the segmentation decoder alone.

\section{CONCLUSION}
\label{sec:con}
In this paper, we propose EndoCaver, a lightweight dual-decoder transformer that jointly performs deblurring and segmentation for endoscopic images. The Global Attention Module enhances encoder features, the Deblurring-Segmentation Aligner transfers restoration cues to segmentation, and the cosine annealing loss scheduler adaptively balances task learning. Experimental results show that EndoCaver achieves robust performance under degradation with low computational cost, enabling practical use in clinical settings.

\clearpage
{
\setlength{\baselineskip}{0.95\baselineskip}
\bibliographystyle{IEEEbib}
\bibliography{strings,refs}

\begin{thebibliography}{10}

\bibitem{morgan2023global}
Eileen Morgan, Melina Arnold, A~Gini, V~Lorenzoni, CJ~Cabasag, Mathieu Laversanne, Jerome Vignat, Jacques Ferlay, Neil Murphy, and Freddie Bray,
\newblock ``Global burden of colorectal cancer in 2020 and 2040: incidence and mortality estimates from globocan,''
\newblock {\em Gut}, vol. 72, no. 2, pp. 338--344, 2023.

\bibitem{munzer2013noisysource}
Bernd M{\"u}nzer, Klaus Schoeffmann, and Laszlo B{\"o}sz{\"o}rmenyi,
\newblock ``Relevance segmentation of laparoscopic videos,''
\newblock in {\em 2013 IEEE international symposium on multimedia}. IEEE, 2013, pp. 84--91.

\bibitem{nazarian2021diagnostic}
Scarlet Nazarian, Ben Glover, Hutan Ashrafian, Ara Darzi, and Julian Teare,
\newblock ``Diagnostic accuracy of artificial intelligence and computer-aided diagnosis for the detection and characterization of colorectal polyps: systematic review and meta-analysis,''
\newblock {\em Journal of medical Internet research}, vol. 23, no. 7, pp. e27370, 2021.

\bibitem{ronneberger2015unet}
Olaf Ronneberger, Philipp Fischer, and Thomas Brox,
\newblock ``U-net: Convolutional networks for biomedical image segmentation,''
\newblock in {\em International Conference on Medical image computing and computer-assisted intervention}. Springer, 2015, pp. 234--241.

\bibitem{xie2021segformer}
Enze Xie, Wenhai Wang, Zhiding Yu, Anima Anandkumar, Jose~M Alvarez, and Ping Luo,
\newblock ``Segformer: Simple and efficient design for semantic segmentation with transformers,''
\newblock {\em Advances in neural information processing systems}, vol. 34, pp. 12077--12090, 2021.

\bibitem{wu2024harmonizing}
Zhuoyu Wu, Qinchen Wu, Wenqi Fang, Wenhui Ou, Quanjun Wang, Linde Zhang, Chao Chen, Zheng Wang, and Heshan Li,
\newblock ``Harmonizing unets: Attention fusion module in cascaded-unets for low-quality oct image fluid segmentation,''
\newblock {\em Computers in Biology and Medicine}, vol. 183, pp. 109223, 2024.

\bibitem{li2025cfformer}
Jiaxuan Li, Qing Xu, Xiangjian He, Ziyu Liu, Daokun Zhang, Ruili Wang, Rong Qu, and Guoping Qiu,
\newblock ``Cfformer: Cross cnn-transformer channel attention and spatial feature fusion for improved segmentation of low-quality medical images,''
\newblock {\em Available at SSRN 5243043}, 2025.

\bibitem{li2024drpvit}
Limiao Li, Keke He, Xiaoyu Zhu, Fangfang Gou, and Jia Wu,
\newblock ``A pathology image segmentation framework based on deblurring and region proxy in medical decision-making system,''
\newblock {\em Biomedical Signal Processing and Control}, vol. 95, pp. 106439, 2024.

\bibitem{dai2024i2u}
Duwei Dai, Caixia Dong, Qingsen Yan, Yongheng Sun, Chunyan Zhang, Zongfang Li, and Songhua Xu,
\newblock ``I2u-net: A dual-path u-net with rich information interaction for medical image segmentation,''
\newblock {\em Medical Image Analysis}, vol. 97, pp. 103241, 2024.

\bibitem{sandler2018mobilenetv2}
Mark Sandler, Andrew Howard, Menglong Zhu, Andrey Zhmoginov, and Liang-Chieh Chen,
\newblock ``Mobilenetv2: Inverted residuals and linear bottlenecks,''
\newblock in {\em Proceedings of the IEEE conference on computer vision and pattern recognition}, 2018, pp. 4510--4520.

\bibitem{mehta2021mobilevit}
Sachin Mehta and Mohammad Rastegari,
\newblock ``Mobilevit: light-weight, general-purpose, and mobile-friendly vision transformer,''
\newblock {\em arXiv preprint arXiv:2110.02178}, 2021.

\bibitem{radford2021clip}
Alec Radford, Jong~Wook Kim, Chris Hallacy, Aditya Ramesh, Gabriel Goh, Sandhini Agarwal, Girish Sastry, Amanda Askell, Pamela Mishkin, Jack Clark, et~al.,
\newblock ``Learning transferable visual models from natural language supervision,''
\newblock in {\em International conference on machine learning}. PmLR, 2021, pp. 8748--8763.

\bibitem{vaswani2017attention}
Ashish Vaswani, Noam Shazeer, Niki Parmar, Jakob Uszkoreit, Llion Jones, Aidan~N Gomez, {\L}ukasz Kaiser, and Illia Polosukhin,
\newblock ``Attention is all you need,''
\newblock {\em Advances in neural information processing systems}, vol. 30, 2017.

\bibitem{xiao2024ctnet}
Bin Xiao, Jinwu Hu, Weisheng Li, Chi-Man Pun, and Xiuli Bi,
\newblock ``Ctnet: Contrastive transformer network for polyp segmentation,''
\newblock {\em IEEE Transactions on Cybernetics}, vol. 54, no. 9, pp. 5040--5053, 2024.

\bibitem{yu2024novel}
Zhenni Yu, Li~Zhao, Tangfei Liao, Xiaoqin Zhang, Geng Chen, and Guobao Xiao,
\newblock ``A novel non-pretrained deep supervision network for polyp segmentation,''
\newblock {\em Pattern Recognition}, vol. 154, pp. 110554, 2024.

\bibitem{jha2019kvasir}
Debesh Jha, Pia~H Smedsrud, Michael~A Riegler, P{\aa}l Halvorsen, Thomas De~Lange, Dag Johansen, and H{\aa}vard~D Johansen,
\newblock ``Kvasir-seg: A segmented polyp dataset,''
\newblock in {\em International conference on multimedia modeling}. Springer, 2019, pp. 451--462.

\bibitem{bernal2015cvcclinic}
Jorge Bernal, F~Javier S{\'a}nchez, Gloria Fern{\'a}ndez-Esparrach, Debora Gil, Cristina Rodr{\'\i}guez, and Fernando Vilari{\~n}o,
\newblock ``Wm-dova maps for accurate polyp highlighting in colonoscopy: Validation vs. saliency maps from physicians,''
\newblock {\em Computerized medical imaging and graphics}, vol. 43, pp. 99--111, 2015.

\bibitem{bernal2012colondb}
Jorge Bernal, Javier S{\'a}nchez, and Fernando Vilarino,
\newblock ``Towards automatic polyp detection with a polyp appearance model,''
\newblock {\em Pattern Recognition}, vol. 45, no. 9, pp. 3166--3182, 2012.

\bibitem{cho2021rethinking}
Sung-Jin Cho, Seo-Won Ji, Jun-Pyo Hong, Seung-Won Jung, and Sung-Jea Ko,
\newblock ``Rethinking coarse-to-fine approach in single image deblurring,''
\newblock in {\em Proceedings of the IEEE/CVF international conference on computer vision}, 2021, pp. 4641--4650.

\bibitem{wu2025rt}
Zhuoyu Wu, Wenhui Ou, Qiawei Zheng, Jiayan Yang, Quanjun Wang, Wenqi Fang, Zheng Wang, Yongkui Yang, and Heshan Li,
\newblock ``Rt-focuser: A real-time lightweight model for edge-side image deblurring,''
\newblock in {\em 2025 IEEE International Conference on Integrated Circuits, Technologies and Applications (ICTA)}. IEEE, 2025, pp. 255--256.

\end{thebibliography}
}

\end{document}